\documentclass[10pt,conference]{IEEEtran}
\usepackage{graphicx}
\usepackage{epsfig}
\usepackage{amsmath,amssymb,amsfonts,cite}

\def\x{{\mathbf x}}
\def\y{{\mathbf y}}

\def\X{{\mathbf X}}

\def\Z{{\mathbf Z}}
\def\Pr{\mathrm{Pr}}
\def\R{{\tilde R}}

\def\w{{\mathbf w}}

\newtheorem{theorem}{Theorem}

\IEEEoverridecommandlockouts

\begin{document}

\title{Multiple Access Channels with Generalized Feedback and Confidential Messages}

%
\author{\authorblockN{Xiaojun Tang\authorrefmark{1},
Ruoheng Liu\authorrefmark{2}, Predrag
Spasojevi\'{c}\authorrefmark{1}, and H. Vincent
Poor\authorrefmark{2}}
\authorblockA{\authorrefmark{1}WINLAB, Rutgers University, North Brunswick, NJ 08902\\
Email: \{xtang, spasojev\}@winlab.rutgers.edu}
\authorblockA{\authorrefmark{2}Princeton University, Princeton, NJ 08544\\
Email: \{rliu, poor\}@princeton.edu}
\thanks{This research was supported by the National Science Foundation under Grants ANI-03-38807 and CNS-06-25637.}
}

\maketitle

\vspace{-0.5cm}
\begin{abstract}

This paper considers the problem of secret communication over a
multiple access channel with generalized feedback. Two trusted users
send independent confidential messages to an intended receiver, in
the presence of a passive eavesdropper. In this setting, an active
cooperation between two trusted users is enabled through using
channel feedback in order to improve the communication efficiency.
Based on rate-splitting and decode-and-forward strategies,
achievable secrecy rate regions are derived for both discrete
memoryless and Gaussian channels. Results show that channel feedback
improves the achievable secrecy rates.
\end{abstract}

\section{Introduction}\label{intro}

The broadcast nature of wireless medium poses both benefits and
penalties for secret communication. The openness of wireless medium
provides opportunities for cooperation between trusted users, which
improves the communication efficiency. On the other hand, it makes
the transmission extremely susceptible to eavesdropping. Anyone
within communication range can listen and possibly extract
information.

Those two opposite aspects are reflected in the system model as
shown in Fig. \ref{channel}, where we consider a multiple access
channel in which two mutually trusted users communicate confidential
messages to an intended receiver, in the presence of a passive
eavesdropper. Channel feedback enables cooperation between two
trusted users and consequently a higher communication efficiency. We
refer to this channel as the \textit{multiple access channel with
generalized feedback and confidential messages} (MAC-GF-CM). The
level of ignorance of the eavesdropper with respect to the
confidential messages is measured by the equivocation rate. This
approach was first introduced by Wyner for the wiretap channel
\cite{Wyner:BSTJ:75}, in which a single source-destination
communication is eavesdropped upon via a degraded channel. Wyner's
formulation was generalized by Csisz{\'{a}}r and K{\"{o}}rner who
determined the capacity region of the broadcast channel with
confidential messages \cite{Csiszar:IT:78}. The Gaussian wiretap
channel was considered in \cite{Leung-Yan-Cheong:IT:78}. More
recently, multi-terminal communication with confidential messages
has been studied further. This work is related to prior works on the
multiple access channel with confidential
messages\cite{Liang:IT:06,Liu:ISIT:06}, the Gaussian multiple access
wiretap channel \cite{Tekin:IT:06}, the interference channel with
confidential messages \cite{Liu:IT:07}, and the relay-eavesdropper
channel \cite{Lai:IT:06,Yusel:CISS:07}.

The multiple access channel with generalized feedback (MAC-GF) without secrecy
consideration was studied in
\cite{King:Stanford:78,Carleial:IT:82,Ozarow:IT:84,Cover:IT:81,Willems:Allerton:83,Willems:IT:85}.
The terminology ``generalized feedback" refers to the wide range of possible
situations, including the MAC without feedback, the MAC with output feedback,
the MAC-GF with independent noise, the MAC with conferencing encoders, the
relay channel and many others. A special case of the Gaussian fading MAC-GF is
the so-called \emph{user cooperation diversity} model proposed in
\cite{Sendonaris:IT:03}.

In this work, we study secret communication over a multiple access
channel with generalized feedback. Based on rate-splitting and
decode-and-forward strategies, achievable secrecy rate regions are
derived for both discrete memoryless and Gaussian MAC-GF-CMs.
Several special cases of the derived achievable secrecy rate region
include the rate regions of the two-user Gaussian multiple access
wiretap channel \cite{Tekin:IT:06}, the relay-eavesdropper channel
\cite{Lai:IT:06,Yusel:CISS:07}, and the MISO wiretap model
\cite{Li:CISS:07}.

The remainder of the paper is organized as follows. Section~\ref{model}
describes the system model. Section \ref{discrete} states our main results on
achievable rate regions for the discrete memoryless MAC-GF-CM. Some
implications of the results are given in Section~\ref{implis}.
Section~\ref{gaussain} states our results for a Gaussian MAC-GF-CM with two
numerical examples.

\section{System Model}\label{model}

\begin{figure}
  \centering
  \includegraphics[width=3in]{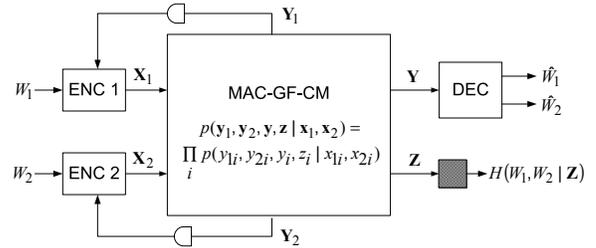}\\
  \caption{The two-transmitter multiple access channel with generalized feedback and confidential messages.}\label{channel}
   \vspace{-0.5cm}
\end{figure}
A two-user multiple access channel with generalized feedback and confidential
messages consists of two transmitters, an intended receiver, and an
eavesdropper, as depicted in Fig.~\ref{channel}. The channel is denoted by
($\mathcal{X}_1 \times \mathcal{X}_2$, $p(y_1,y_2,y,z|x_1,x_2)$, $\mathcal{Y}_1
\times \mathcal{Y}_2 \times \mathcal{Y} \times \mathcal{Z}$), where
$\mathcal{X}_1$ and $\mathcal{X}_2$ are input alphabets; $\mathcal{Y}$ and
$\mathcal{Z}$ are output alphabets at the intended receiver and the
eavesdropper, respectively; $\mathcal{Y}_1$ and $\mathcal{Y}_2$ are the
feedback channel output alphabets; and $p(y_1,y_2,y,z|x_1,x_2)$ is the
transition probability matrix. The channel is memoryless and time-invariant in
the sense that
\begin{equation*}\label{dm}
    p(y_{1i},y_{2i},y_{i},z_{i}|\x_1^i,\x_2^i,\y_1^{i-1},\y_2^{i-1})=p(y_{1i},y_{2i},y_{i},z_{i}|x_{1i},x_{2i})
\end{equation*}
where $\x_t^i=[x_{t1},x_{t2},\dots,x_{ti}]$ for $t=1,2$. The
superscript will be dropped when $i=n$ in order to simplify
notations.

Encoder 1 and encoder 2 send independent messages $W_1 \in
\mathcal{W}_1 = \{1,\dots, M_1\}$ and $W_2 \in \mathcal{W}_2 =
\{1,\dots, M_2\}$ to the intended receiver in $n$ channel uses, in a
cooperative way by using the feedback signals $(\y_1,\y_2)$. For
$t=1,2$, a stochastic encoder $f_t$ for user $t$ is specified by a
matrix of conditional probabilities $f(x_{ti}|w_t,\y_t^{i-1})$,
where $x_{ti} \in \mathcal{X}_t$, $w_t \in \mathcal{W}_t$,
$\y_t^{i-1} \in \mathcal{Y}_t^{i-1}$ and
$\sum_{x_{ti}}f(x_{ti}|w_t,\y_t^{i-1})=1$,
for $i=1,\dots, n$, where $f(x_{ti}|w_t,\y_t^{i-1})$ is the
probability that encoder $t$ outputs $x_{ti}$ when message $w_t$ is
being sent and $\y_t^{i-1}$ has been observed at encoder $t$.

The decoder uses the output sequence $y^n$ to compute its estimate
$(\hat{w}_1,\hat{w}_2)$ of $(w_1,w_2)$. The decoding function is
specified by a mapping $\phi: \mathcal{Y}^n \rightarrow
\mathcal{W}_1 \times \mathcal{W}_2$.

An ($M_1,M_2,n,P_e$) code for the MAC with generalized feedback and
confidential messages consists of two sets of $n$ encoding functions
${f_{ti}}$, $t=1,2$, $i=1,\dots,n$ and a decoding function $\phi$ so
that its average probability of error is

\begin{equation}\label{pe}
    P_e=\sum_{(w_1,w_2)}\frac{1}{M_1M_2}\Pr\left\{\phi(\y) \neq (w_1,w_2) | (w_1,w_2)
    \mbox{sent}\right\}.
\end{equation}

The level of ignorance of the eavesdropper with respect to the confidential
messages is measured by the equivocation rate $H(W_1,W_2|\Z)/n$.

A rate pair $(R_1,R_2)$ is achievable for the MAC with generalized feedback and
confidential messages if, for any $\epsilon>0$, there exists an
($M_1,M_2,n,P_e$) code so that

\begin{equation}\label{ach_def1}
    M_1 \geq 2^{nR_1},~ M_2 \geq 2^{nR_2}, ~ P_e \leq \epsilon
\end{equation}
\begin{equation}\label{ach_def2}
\text{and} \qquad  R_1 + R_2 - H(W_1,W_2|\Z)/n \leq \epsilon \quad
\qquad ~
\end{equation}
for all sufficiently large $n$. The secrecy capacity region is the closure of
the set of all achievable rate pairs $(R_1,R_2)$.

We note that the perfect secrecy condition (\ref{ach_def2}) implies
\begin{equation}\label{ach_def3}
    R_1 -\frac{1}{n}H(W_1|\Z) \leq \epsilon~~\text{and}~~R_2
-\frac{1}{n}H(W_2|\Z) \leq \epsilon.
\end{equation}
and therefore the \textit{joint} perfect secrecy requirement is
stronger than the \textit{individual} perfect secrecy requirement.

This can be shown as follows:
\begin{eqnarray*}
 H(W_1|\Z) &=& H(W_1,W_2|\Z)-H(W_2|W_1,\Z) \\
   &\geq& H(W_1) + H(W_2) - n\epsilon - H(W_2|W_1,\Z) \\
   &\ge& H(W_1) - n\epsilon \\
   &= &n(R_1 -\epsilon).
\end{eqnarray*}
Similarly, we can show that (\ref{ach_def2}) implies $H(W_2|\Z) \geq
n(R_2 -\epsilon)$.

\section{Discrete Memoryless Channels}\label{discrete}

We first state our results for discrete memoryless channels.

\begin{theorem}\label{TH1} (Partial Decode-and-Forward)

For a discrete memoryless MAC with generalized feedback and
confidential messages, the secrecy rate region
$\mathcal{R}(\pi_{I})$ is achievable, where $\mathcal{R}(\pi_{I})$
is the closure of the convex hull of all $(R_1,R_2)$ satisfying
\begin{align}
 \left\{ \begin{array}{ll}
  R_1 = R_{10}+R_{12}, R_2 = R_{20}+R_{21}:\\
  R_{10} + \R_{10} \leq I(X_1;Y|X_2,V_1,U),\\
  R_{20} + \R_{20}\leq I(X_2;Y|X_1,V_2,U),\\
  R_{10} + R_{20} + \R_{10}+ \R_{20} \leq I(X_1,X_2;Y|V_1,V_2,U),\\
  R_{12} + \R_{12} \leq I(V_1; Y_2|X_2,U),\\
  R_{21} + \R_{21} \leq I(V_2; Y_1|X_1,U),\\
  R_{10} + R_{20} + R_{12} + R_{21} \\ \qquad \qquad \qquad \qquad \leq I(X_1,X_2;Y) - I(X_1,X_2;Z).\\
  R_{10},R_{20},R_{12},R_{21}\geq
  0,\\
 (\R_{10},\R_{20},\R_{12},\R_{21}) \in \mathcal{C}(\R_{10},\R_{20},\R_{12},\R_{21})
       \end{array} \right.
\end{align}
where
\begin{align}
 \mathcal{C}(&\R_{10},\R_{20},\R_{12},\R_{21}) = \notag\\
 &\left\{ \begin{array}{ll}
 (\R_{10}, \R_{20},\R_{12},\R_{21} \geq 0):\\
  \R_{10} \leq I(X_1;Z|X_2,V_1,U),\\
  \R_{20} \leq I(X_2;Z|X_1,V_2,U),\\
  \R_{10} + \R_{20} \leq I(X_1,X_2;Z|V_1,V_2,U),\\
  \R_{10} + \R_{20} + \R_{21} + \R_{12} = I(X_1,X_2;Z).\\
  \end{array} \right.
\end{align}
and $\pi_{I}$ denotes the class of joint probability mass functions
$p(u,v_1,v_2,x_1,x_2,y_1,y_2,y,z)$ that factor as
\begin{equation*}
    p(u)p(v_1,x_1|u)p(v_2,x_2|u)p(y_1,y_2,y,z|x_1,x_2).
\end{equation*}

\end{theorem}


Theorem \ref{TH1} illustrates a rate-splitting strategy. The rates $R_1$ and
$R_2$ are split as $R_1= R_{10} + R_{12}$ and $R_2= R_{20} + R_{21}$, where
$R_{12}$ and $R_{21}$ are the rates of information sent by both transmitters
cooperatively to the intended receiver, while $R_{10}$ and $R_{20}$ are the
rates of non-cooperative information sent by user 1 and user 2 individually to
the receiver. The random variable $U$ represents cooperative resolution
information sent by both transmitters. $V_1$ represents information (at rate
$R_{12}$) that user 1 sends to user 2 to enable cooperation. $V_2$ represents
information (at rate $R_{21}$) that user 2 sends to user 1 to enable
cooperation.

$\R_{10}$, $\R_{20}$, $\R_{12}$ and $\R_{21}$ represent the rates sacrificed in
order to confuse the eavesdropper completely. The sum rate loss is
$I(X_1,X_2;Z)$. When we set $Z = \varnothing $ (in the case of no
eavesdropper), $\R_{10}=\R_{20}=\R_{12}=\R_{21}=0$, and hence, our result
becomes the rate region of the MAC with general feedback as given in
\cite{Willems:Allerton:83}.

The achievability scheme is based on the combination of
superposition block Markov encoding \cite{Cover:IT:81}, backward
decoding \cite{Willems:IT:85} and random binning
\cite{Wyner:BSTJ:75,Slepian:IT:73}. We outline the proof in the
Appendix.

\textit{Remark 1}: The rate region may be enlarged by using the channel
prefixing technique in \cite[Lemma~4]{Csiszar:IT:78}. However, we do not follow
this approach in this paper to avoid its complicated notation and the
intractability of its evaluation.

If we require that $R_{1}= R_{12}$ and $R_2=R_{21}$, that is, all
information is sent cooperatively and each user can fully decode the
other user's message, we have the following result.

\begin{theorem}(Full Decode-and-Forward)\label{TH2}

The secrecy rate region $\mathcal{R}(\pi_{II})$ is achievable, where
$\mathcal{R}(\pi_{II})$ is the closure of the convex hull of all
$(R_1,R_2)$ satisfying
\begin{align}
    \left\{ \begin{array}{ll}
 (R_1,R_2 \geq 0):\\
  R_1 \leq I(X_1; Y_2|X_2,U),\\
  R_2 \leq I(X_2; Y_1|X_1,U),\\
  R_1 + R_2 \leq \min \{I(X_1; Y_2|X_2,U) \\ \qquad \qquad \quad
  + I(X_2; Y_1|X_1,U),~I(X_1,X_2;Y)\}\\
  \qquad \qquad \quad - I(X_1,X_2;Z).
        \end{array} \right.
\end{align}
where $\pi_{II}$ denotes the class of joint probability mass
functions $p(u,x_1,x_2,y_1,y_2,y,z)$ that factor as
\begin{equation*}
    p(u)p(x_1|u)p(x_2|u)p(y_1,y_2,y,z|x_1,x_2).
\end{equation*}

\end{theorem}

\section{Some Implications of the Results}\label{implis}

Next, we discuss some implications of our main result. We consider
several special cases of Theorems~\ref{TH1} and~\ref{TH2}, which are
consistent with the recent results in
\cite{Tekin:IT:06,Lai:IT:06,Yusel:CISS:07,Li:CISS:07}.

\subsection{Multiple Access Wiretap Channel}

An achievable rate region for the Gaussian multiple access wiretap channel is
given in \cite{Tekin:IT:06}, which is the special case when neither user can
obtain feedback, i.e., $Y_1= \varnothing $ and $Y_2= \varnothing $. We set
$V_1=V_2=U= \varnothing$ in Theorem \ref{TH1} and have the achievable region
$\mathcal{R}(\pi_{MAC-WT})$, which is the closure of the convex hull of all
$(R_1,R_2)$ satisfying
\begin{align}
    \left\{ \begin{array}{ll}
 (R_1,R_2 \geq 0):\\
  R_1 \leq I(X_1; Y|X_2) - I(X_1; Z),\\
  R_2 \leq I(X_2; Y|X_1) - I(X_2; Z),\\
  R_1 + R_2 \leq I(X_1,X_2;Y) - I(X_1,X_2;Z),\\
        \end{array} \right.
\end{align}
where $\pi_{MAC-WT}$ is the class of all distributions that factor as
$p(x_1,x_2,y,z)= p(x_1)p(x_2)p(y,z|x_1,x_2)$.

\subsection{Relay-Eavesdropper Channel}

An achievable rate region for the relay-eavesdropper channel is given in
\cite{Lai:IT:06,Yusel:CISS:07}, which is the case when only user 1 has
confidential messages to send and user 2 is a relay to help with the
decode-and-forward strategy; therefore $R_2=0$ and $Y_1= \varnothing$. We set
$V_2=\varnothing$ and $U=X_2$ in Theorem \ref{TH2} and the achievable rate
satisfies
\begin{equation}
  R_1 \leq [\min\{I(X_1;Y_2|X_2),I(X_1,X_2;Y)\}- I(X_1,X_2;Z)]^{+},
\end{equation}
for all distributions that factor as $p(x_1,x_2,y_2,y,z) =
p(x_1,x_2)p(y_2,y,z|x_1,x_2)$. This result is consistent with
\cite[Theorem~2]{Lai:IT:06}.

\subsection{MISO Wiretap Channel}

When each transmitter can obtain perfect channel feedback, i.e., $Y_2=V_1$ and
$Y_1=V_2$, we have a virtual MISO wiretap channel. We set $V_1=X_1$ and
$V_2=X_2$ in Theorem \ref{TH1}. The achievable secrecy rate of the MISO channel
is given by
\begin{align}
    \begin{array}{ll}
   R = R_1 + R_2 \leq [I(X_1,X_2;Y)- I(X_1,X_2;Z)]^{+},
        \end{array}
\end{align}
for all distributions that factor as $p(x_1,x_2,y,z) = p(x_1,x_2)
p(y,z|x_1,x_2)$. This result is consistent with \cite{Li:CISS:07}.

\section{Gaussian Channels}\label{gaussain}

\begin{figure}
  \centering
  \includegraphics[width=3in]{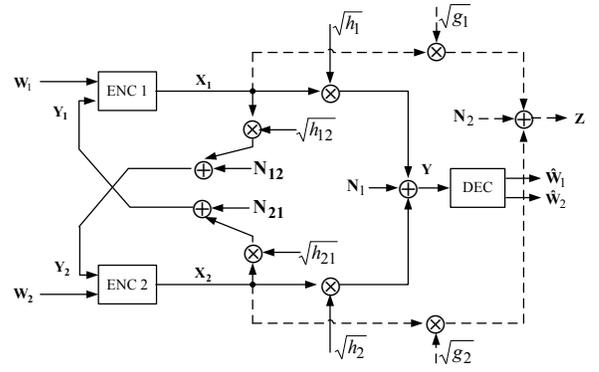}\\
  \caption{A Gaussian MAC-GF with confidential messages}\label{channel2}
  \vspace{-0.5cm}
\end{figure}

In this section, we consider a Gaussian MAC-GF-CM, as depicted in
Fig. \ref{channel2}. Each mutually trusted user receives an
attenuated and noisy version of the partner's signal and uses that
signal, in conjunction with its own message, to construct the
transmit signal. The intended receiver and a passive eavesdropper
each get a noisy version of the sum of the attenuated signals of
both users. The signal model is therefore
\begin{eqnarray}
  Y_1 &=& \sqrt{h_{21}}X_{2} + N_{21} \notag\\
  Y_2 &=& \sqrt{h_{12}}X_{1} + N_{12} \notag\\
  Y &=& \sqrt{h_{1}}X_{1} + \sqrt{h_{2}}X_{2} + N_1 \notag\\
  Z &=& \sqrt{g_{1}}X_{1} + \sqrt{g_{2}}X_{2} + N_2.
\end{eqnarray}
where $h_i,g_i$ for $i=1,2$ are main and eavesdropper channel gains
respectively; $h_{12}$ and $h_{21}$ are feedback channel gains, as
shown in Fig. \ref{channel2}.
We assume the following: the transmitted signal $X_t$ has an average
power constraint
\begin{equation}
    \frac{1}{n}\sum_{i=1}^{n}E[X_{ti}^{2}] \leq P_t, ~~t=1,2;
\end{equation}
and the noise terms $N_1$, $N_2$, $N_{12}$, and $N_{21}$ are
independent white zero-mean unit-variance complex Gaussian, i.e.,
$N_1 \sim \mathcal{N}(0,1)$, $N_2 \sim \mathcal{N}(0,1)$, $N_{12}
\sim \mathcal{N}(0,1)$, and $N_{21} \sim \mathcal{N}(0,1)$.

Let $V_1$, $V_2$, $X_1$, and $X_2$ be jointly Gaussian with
\begin{eqnarray}
  V_1 &=& \sqrt{P_{U1}}U + \sqrt{P_{12}}U'_1 \notag\\
  V_2 &=& \sqrt{P_{U2}}U + \sqrt{P_{21}}U'_2 \notag\\
  X_1 &=& V_1 + \sqrt{P_{10}}U''_1 \notag\\
  X_2 &=& V_{2} + \sqrt{P_{20}}U''_2
\end{eqnarray}
where $U$, $U'_1$, $U'_2$, $U''_1$, and $U''_2$ are independent zero
mean unit variance Gaussian. The terms $P_{U1}$, $P_{12}$, $P_{10}$,
$P_{U2}$, $P_{21}$ and $P_{20}$ denote the corresponding power
allocation, where
\begin{align}
P_1=P_{U1}+P_{12}+P_{10}~~\text{and}~~P_2=P_{U2}+P_{21}+P_{20}.
\end{align}

Following the achievability proof for the discrete memoryless
channel, we have the following result for the Gaussian multiple
access channel with feedback.

\begin{theorem}(Partial Decode-and-Forward) \label{TH3}

An achievable secrecy rate region $\mathcal{R}_{G}^{I}$ is the closure of the
convex hull of all rate pairs $(R_1,R_2)$ with
\begin{align}
 \left\{ \begin{array}{ll}
  R_1 = R_{10} + R_{12}, R_2 = R_{20}+R_{21}:\\
  R_{10} + \R_{10} \leq C(h_1P_{10}),\\
  R_{20} + \R_{20} \leq C(h_2P_{20}),\\
  R_{10} + R_{20} + \R_{10}+ \R_{20} \leq C(h_1P_{10}+h_2P_{20}),\\
  R_{12} + \R_{12} \leq C(\frac{h_{12}P_{12}}{1+h_{12}P_{10}}),\\
  R_{21} + \R_{21} \leq C(\frac{h_{21}P_{21}}{1+h_{21}P_{20}}),\\
  R_{10} + R_{20} + R_{12} + R_{21} \leq \\ \quad \quad  C\left(h_1P_1+h_2P_2+2\sqrt{h_1h_2P_{U1}P_{U2}}\right)\\\quad\quad -C\left(g_1P_1+g_2P_2+2\sqrt{g_1g_2P_{U1}P_{U2}}\right).\\
  R_{10},R_{20},R_{12},R_{21} \geq 0,\\
 (\R_{10},\R_{20},\R_{12},\R_{21}) \in \mathcal{C}(\R_{10},\R_{20},\R_{12},\R_{21})
       \end{array} \right.
\end{align}
where\begin{align}
 \mathcal{C}(&\R_{10},\R_{20},\R_{12},\R_{21}) = \notag\\
 &\left\{ \begin{array}{ll}
 (\R_{10}, \R_{20},\R_{12},\R_{21} \geq 0):\\
  \R_{10} \leq C(g_1P_{10}),\\
  \R_{20} \leq C(g_2P_{20}),\\
  \R_{10} + \R_{20} \leq C(g_1P_{10}+g_2P_{20}),\\
  \R_{10} + \R_{20} + \R_{21} + \R_{12} = \\
  \qquad \quad C\left(g_1P_1+g_2P_2+2\sqrt{g_1g_2P_{U1}P_{U2}}\right),
  \end{array} \right.
\end{align}
and $C(x)\triangleq(1/2)\log(1+x)$.
\end{theorem}

\begin{figure}
  \centering
  \includegraphics[width=3.2in]{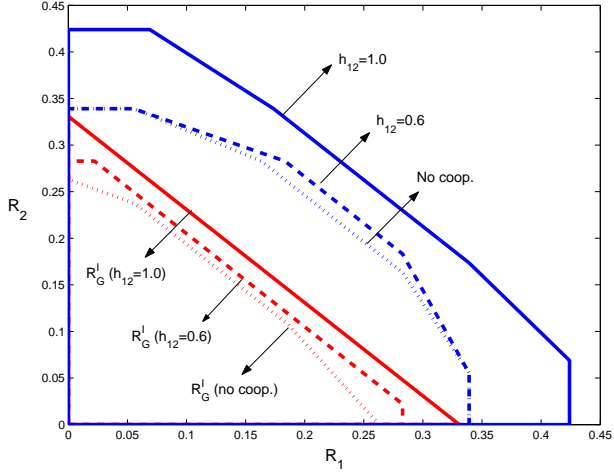}\\
  \caption{Regular Rate regions and secrecy rate regions $\mathcal{R}_{G}^{I}$ for
  $h_1=0.6$, $h_2=0.6$, $g_1=0.2$, $g_2=0.1$, $P_1=1$, $P_2=1$ under different cooperation
  conditions $h_{12}=h_{21}\in[0, 0.6, 1.0]$, where $h_{12}=h_{21}=0$ means no cooperation.}\label{cmac1}
\end{figure}

As a numerical example, we show in Fig. \ref{cmac1} the ``regular"
rate region (without the secrecy constraint) and the secrecy rate
region $\mathcal{R}_{G}^{I}$ for $h_1=0.6$, $h_2=0.6$, $g_1=0.2$,
$g_2=0.1$, $P_1=1$ and $P_2=1$ under different cooperation
conditions $h_{12}=h_{21}\in[0, 0.6, 1.0]$. When $h_{12}=h_{21}=0$,
there is no cooperation between the two encoders, which corresponds
to the multiple access wiretap channel. Both the regular rate region
and the secrecy rate region are significantly enlarged when the
channel gains between the two users ($h_{21}$ and $h_{12}$) become
larger, which shows the benefits due to cooperation. Comparing with
the regular rate region, the secrecy rate region suffers rate loss
due to the secrecy constraint and furthermore, the secrecy rate
region is increasingly dominated by the sum rate constraint, as
depicted in Fig. \ref{cmac1}.

Next, we give the secrecy rate region when each user can fully
decode the message sent by the other user.

\begin{theorem}(Full Decode-and-Forward) \label{TH4}

An achievable secrecy rate region $\mathcal{R}_{G}^{II}$ is the closure of the
convex hull of all rate pairs $(R_1,R_2)$ with
\begin{align}
    \left\{ \begin{array}{ll}
 (R_1,R_2 \geq 0):\\
  R_1 \leq C(h_{12}P_{12}),\\
  R_2 \leq C(h_{21}P_{21}),\\
  R_1 + R_2 \leq \min \{C(h_{12}P_{12}) + C(h_{21}P_{21}),
  \\ \quad \quad \quad C\left(h_1P_1+h_2P_2+2\sqrt{h_1h_2P_{U1}P_{U2}}\right)\}
  \\ \quad \quad\quad  -C\left(g_1P_1+g_2P_2+2\sqrt{g_1g_2P_{U1}P_{U2}}\right).
        \end{array} \right.
\end{align}
\end{theorem}

We summarize the secrecy sum rates of partial and full decode-and-forward
strategies in the following theorem.
\begin{theorem} (Sum Rate) \label{TH5}
The maximal achievable sum rate in $\mathcal{R}_{G}^{I}$ is
\begin{align}\label{sum_rate1}
R^{I}&= \min \Bigl\{ C\left(h_1P_1+h_2P_2+2\sqrt{h_1h_2P_{U1}P_{U2}}\right),\nonumber\\
& \qquad C\left(\frac{h_{12}P_{12}}{1+h_{12}P_{10}}\right) + C\left(\frac{h_{21}P_{21}}{1+h_{21}P_{20}}\right)\nonumber \\
& \quad +C\left(h_1P_{10}+h_2P_{20}\right)\Bigr\}\nonumber\\
& \quad -C\left(g_1P_1+g_2P_2+2\sqrt{g_1g_2P_{U1}P_{U2}}\right);
\end{align}
the maximum achievable sum rate in $\mathcal{R}_{G}^{II}$ is
\begin{align}\label{sum_rate2}
   R^{II}&= \min \Bigl\{C\left(h_1P_1+h_2P_2+2\sqrt{h_1h_2P_{U1}P_{U2}}\right),\nonumber\\
   & \qquad  C(h_{12}P_{12}) + C(h_{21}P_{21})\Bigr\}\nonumber\\
   & \quad -C\left(g_1P_1+g_2P_2+2\sqrt{g_1g_2P_{U1}P_{U2}}\right).
\end{align}
Furthermore, $R^{I}=R^{II}$ when $h_{12} \geq h_1$ and $h_{21} \geq h_2.$
\end{theorem}

The proof of Theorem~\ref{TH5} is provided in the Appendix.

In Fig. \ref{cmac2}, we illustrate secrecy rate regions
$\mathcal{R}_{G}^{I}$ and $\mathcal{R}_{G}^{II}$ for $h_1=0.6$,
$h_2=0.6$, $g_1=0.2$, $g_2=0.1$, $P_1=1$ and $P_2=1$ under different
cooperation conditions $h_{12}=h_{21}\in[0.2, 0.55, 1.0]$. Comparing
with $\mathcal{R}_{G}^{I}$, $\mathcal{R}_{G}^{II}$ suffers a
significant rate loss when $h_{12}$ and $h_{21}$ are small
($h_{12}=h_{21}=0.2$) as expected. When $h_{12}$ and $h_{21}$
increase, the rate loss is reduced. When $h_{12}$ and $h_{21}$ are
large enough, $\mathcal{R}_{G}^{II}$ and $\mathcal{R}_{G}^{I}$
coincide. This observation is partially verified by
Theorem~\ref{TH5}.

\begin{figure}
  \centering
  \includegraphics[width=3.2in]{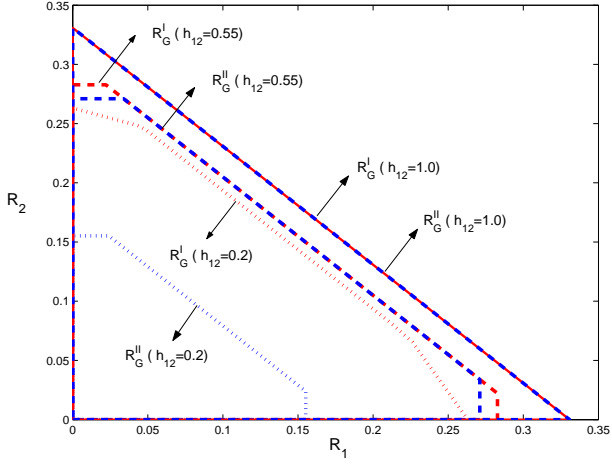}\\
  \caption{Secrecy rate regions $\mathcal{R}_{G}^{I}$ and $\mathcal{R}_{G}^{II}$ for
  $h_1=0.6$, $h_2=0.6$, $g_1=0.2$, $g_2=0.1$, $P_1=1$, $P_2=1$ under different cooperation
  conditions: $h_{12}=h_{21}\in[0.2, 0.55, 1.0]$. }\label{cmac2}
  \vspace{-0.5cm}
\end{figure}

\appendix
\noindent \textit{Proof:~}(\textbf{Theorem 1}) The transmission is
performed for $B+1$ blocks of length $n_1$, where both $B$ and $n_1$
are sufficiently large and $n=(B+1)n_1$.

The random code generation is described as follows.

We fix $p(u)$, $p(v_1,x_1|u)$ and $p(v_2,x_2|u_2)$ and split the
rate pair $(R_1,R_2)$ as $R_1 = R_{12}+R_{10}$ and $R_2 =
R_{21}+R_{20}$. Let
\begin{equation}\label{RB}
    \R_{12}+ \R_{10} + \R_{21} + \R_{20} = I(X_1,X_2;Z) - \epsilon_1
\end{equation}
where $\epsilon_1 >0$ and $\epsilon_1 \rightarrow 0$ as $n_1 \rightarrow
\infty$. Let $R'_1= R_{12} + \R_{12}$, $R''_1= R_{10} + \R_{10}$, $R'_2= R_{21}
+ \R_{21}$ and $R''_2= R_{20} + \R_{20}$.

\noindent\textbf{Code Construction}:
\begin{itemize}[\setlabelwidth{Z}]

\item

Generate $2^{n_1(R'_1+R'_2)}$ codewords $u^{n_1}(\w_0)$ by choosing
the $u_i(\w_0)$ independently according to $p(u)$ for
$i=1,2,\dots,n_1$, where $\w_0=1,2,\dots,2^{n_1(R'_1+R'_2)}$.

\item

For each $\w_0$, generate $2^{n_1R'_1}$ codewords
$v_1^{n_1}(\w_0,\w'_1)$ by choosing the $v_{1i}(\w_0,\w'_1)$
independently according to $p(v_1|u)$ for $i=1,2,\dots,n_1$, where
$\w'_1=1,2,\dots,2^{n_1R'_1}$.

\item

For each tuple $(\w_0,\w'_1)$, generate $2^{n_1R''_1}$ codewords
$x_1^{n_1}(\w_0,\w'_1,\w''_1)$ by choosing the
$x_{1i}(\w_0,\w'_1,\w''_1)$ independently according to
$p(x_1|u,v_1)$ for $i=1,2,\dots,n_1$, where
$\w''_1=1,2,\dots,2^{n_1R''_1}$.

\end{itemize}

The codebooks for user 2 are generated in the same way, except that there are
$2^{n_1R'_2}$ and $2^{n_1R''_2}$ codewords in each of the $v_2^{n_1}$ and
$x_2^{n_1}$ codebooks, respectively. The same codebooks will be used for all
$B+1$ blocks during the encoding.

\noindent\textbf{Encoding}: \quad Message $w_1$ has $n_1(R_1B+R_{10})$ bits and
is split into two parts: $w'_1$ with $n_1R_{12}B$ bits and $w''_1$ with
$n_1R_{10}(B+1)$ bits, respectively. Message $w_2$ is similarly divided into
$w'_2$ and $w''_2$. Each of the four messages $w'_1$, $w''_1$, $w'_2$ and
$w''_2$ is further divided into $B$ sub-blocks of equal lengths for each
message. They are denoted by $w'_{1,b}$, $w''_{1,b}$, $w'_{2,b}$ and
$w''_{2,b}$, respectively, for $b=1,2,\dots,B+1$. Let
\begin{equation}\label{ran_bin}
    \w'_{1,b}= (w'_{1,b},\tilde{w}'_{1,b}), ~\text{and}~ \w''_{1,b}=
    (w''_{1,b},\tilde{w}''_{1,b}),
\end{equation}
where $\tilde{w}'_{1,b}$ and $\tilde{w}''_{1,b}$ are uniformly and
independently chosen at random from $\{1,2,\dots, 2^{n_1\R_{12}}\}$ and
$\{1,2,\dots, 2^{n_1\R_{10}}\}$ respectively. We also choose $\w'_{1,0} =
(1,1)$ and $\w'_{1,B+1} = (1,1)$. The $\w'_{2,b}$ and $\w''_{2,b}$ for
$b=1,\dots,B+1$ are formed in the same way.

Suppose that encoder 1 has obtained $\w'_{2,b-1}$ and encoder 2 has obtained
$\w'_{1,b-1}$ before block $b$. By forming
$\w_{0,b}=(\w'_{1,b-1},\w'_{2,b-1})$, encoder 1 transmits
$x_1^n(\w_{0,b},\w'_{1,b},\w''_{1,b})$; encoder 2 transmits
$x_2^n(\w_{0,b},\w'_{2,b},\w''_{2,b})$ in block $b$.

\noindent\textbf{Decoding}:\quad  All decodings are based on the typical set
decoding. After the transmission of block $b$ is completed, user 1 has seen
$y_{1,b}^{n_1}$. It tries to decode $\w'_{2,b}$. User 2 operates in the same
way.

The intended receiver waits until all $B+1$ blocks of transmission are
completed and performs backward decoding. Given $y_{B+1}^{n_1}$, it tries to
decode $(\w_{B+1},\w''_{1,B+1},\w''_{2,B+1})$. Assuming that the decoding for
block $B+1$ is correct, the decoder next considers $y_B^{n_1}$ to decode
$(\w_{B},\w''_{1,B},\w''_{2,B})$. The decoder continues until it decodes all
blocks.

\noindent\textbf{Error Analysis}: \quad Following similar steps to the error
analysis for the MAC-GF in \cite{Willems:Allerton:83}, we found that the
intended receiver can decode all $\w'_{b}, \w''_{b}$ and therefore $w_1, w_2$
with error probability less than any $\epsilon >0$ if
\begin{align*}
R_{21} + \R_{21} = R'_2  &\leq I(V_2; Y_1|X_1,U),\\
R_{12} + \R_{12} = R'_1  &\leq I(V_1; Y_2|X_2,U),\\
R_{10} + \R_{10} = R''_1 &\leq I(X_1;Y|X_2,V_1,U),\\
R_{20} + \R_{20} = R''_2 &\leq I(X_2;Y|X_1,V_2,U),\\
R_{10} + R_{20} + \R_{10}+ \R_{20} &\leq I(X_1,X_2;Y|V_1,V_2,U),
\end{align*}
and
\begin{align*}
\qquad R_{10} + R_{20} + R_{12} + R_{21}  &\leq I(X_1,X_2;Y)-I(X_1,X_2;Z),
\end{align*}
for sufficiently large $n_1$, where we also used (\ref{RB}).

\noindent\textbf{Equivocation}:\quad Now we consider the equivocation,
\begin{align}\label{equiv}
 H(&W_1,W_2|\Z) \nonumber \\
   &= H(W_1,W_2,\Z) - H(\Z)\nonumber\\
   &= H(W_1,W_2,\Z,\X_1,\X_2) - H(\X_1,\X_2|W_1,W_2,\Z) -H(\Z)\nonumber\\
   &= H(\X_1,\X_2) + H(W_1,W_2,\Z | \X_1,\X_2) - H(Z) \nonumber\\ & \quad  - H(\X_1,\X_2|W_1,W_2,\Z) \nonumber\\
   &\geq H(\X_1,\X_2) + H(\Z |\X_1,\X_2) -H(\Z)\nonumber\\ & \quad - H(\X_1,\X_2|W_1,W_2,\Z)\nonumber\\
   &= H(\X_1,\X_2) - I(\X_1,\X_2;\Z) -
   H(\X_1,\X_2|W_1,W_2,\Z),
\end{align}
and we can bound each term in the above.

The first term in (\ref{equiv}) is given by
\begin{align}\label{equiv_T1}
 H(&\X_1,\X_2) \nonumber \\
   &= n_1B(R_{10}+R_{20}+R_{12}+R_{21})+n_1(R_{10}+R_{20})\nonumber\\
   &\quad + n_1B(\R_{10}+\R_{20}+\R_{12}+\R_{21})+n_1(\R_{10}+\R_{20})\nonumber\\
   &\geq n_1B(R_1+R_2)+ n_1B\left[I(X_1,X_2;Z)-\epsilon_1\right].
\end{align}

Since the channel is memoryless, the second term in (\ref{equiv})
can be bounded by
\begin{equation}\label{equiv_T2}
  I(\X_1,\X_2;\Z) \leq n_1(B+1)\left[I(X_1,X_2;Z)-\delta_1\right]
\end{equation}
where $\delta_1 \rightarrow 0$ as $n_1 \rightarrow \infty$.

We next show that the third term can be bounded by
\begin{equation}\label{equiv_T3}
    H(\X_1,\X_2|W_1,W_2,\Z) \leq n_1(B+1)\delta_2.
\end{equation}
In order to calculate $H(\X_1,\X_2|W_1,W_2,\Z)$, we consider the
following situation: the transmitters send fixed messages $W_1 =
w_1, W_2=w_2$. Now, the eavesdropper also performs backward decoding
to decode all $(\w_{0,b},\w''_{1,b}$ and $\w''_{2,b})$. We can show
that the error probability is less than any $\epsilon>0$ if
\begin{align}
  \R_{10} &\leq I(X_1;Z|X_2,V_1,U),\label{macZ1}\\
  \R_{20} &\leq I(X_2;Z|X_1,V_2,U),\label{macZ2}\\
\text{and} \qquad \R_{10} + \R_{20} &\leq I(X_1,X_2;Z|V_1,V_2,U)\label{macZ3},
\end{align}
for sufficiently large $n_1$. In other words, given message $(w_1,w_2)$, the
eavesdropper can decode $(\X_1,\X_2)$ under conditions (\ref{macZ1}),
(\ref{macZ2}) and (\ref{macZ3}). Therefore, Fano's inequality implies that
\begin{equation}\label{fano}
    H(\X_1,\X_2|W_1=w_1,W_2=w_2,\Z) \leq n_1(B+1)\delta_2.
\end{equation}
Hence,
\begin{eqnarray*}
 \lefteqn{ H(\X_1,\X_2|W_1,W_2,\Z)} \\
   &=& \sum_{w_1}\sum_{w_2}p(W_1=w_1)p(W_2=w_2)\\&&{}H(\X_1,\X_2|W_1=w_1,W_2=w_2,\Z)\\
   &\leq& n_1(B+1)\delta_2.
\end{eqnarray*}
By using (\ref{equiv_T1}), (\ref{equiv_T2}) and (\ref{equiv_T3}), we
can rewrite (\ref{equiv}) as
\begin{eqnarray*}
 \lefteqn{ H(W_1,W_2|\Z)} \\
   &\geq& n_1B(R_1+R_2)+ n_1B\left[I(X_1,X_2;Z)-\epsilon_1\right]\\
   &&{}-n_1(B+1)\left[I(X_1,X_2;Z)-\delta_1\right]-n_1(B+1)\delta_2\\
   &\geq& n_1B(R_1+R_2)-n_1I(X_1,X_2;Z)-n(B+1)\epsilon.
\end{eqnarray*}
The equivocation rate is therefore
\begin{eqnarray*}
 \lefteqn{\frac{1}{n} H(W_1,W_2|\Z) = \frac{1}{n_1(B+1)} H(W_1,W_2|\Z)} \\
   &\geq& (1-\frac{1}{B+1})(R_1+R_2)- \frac{1}{B+1}I(X_1,X_2;Z)-\epsilon.
\end{eqnarray*}

For sufficiently large $B$, we have
\begin{equation}\label{equiv_rate}
\frac{1}{n} H(W_1,W_2|\Z) \geq R_1 + R_2 -\epsilon,
\end{equation}
which is the perfect secrecy requirement defined by
(\ref{ach_def2}).

\bigskip
\noindent \textit{Proof:~}(\textbf{Theorem 5}) The sum rates
(\ref{sum_rate1}) and (\ref{sum_rate2}) can be derived based on
Theorems~\ref{TH3} and~\ref{TH4}, respectively. Hence, we need only
to show that $P_{10}=P_{20}=0$ in (\ref{sum_rate1}) is optimal to
maximize $R^{I}$, when $h_{12} \geq h_1$ and $h_{21} \geq h_2$.

It is easy to show that $R^{I}$ can be written as
\begin{equation*}
    R^{I} = \frac{1}{2}\min \{\log(T_1),\log(T_2)+\log(T_3)\},
\end{equation*}
where
\begin{align*}
    T_1&=\frac{1+h_1P_1+h_2P_2+2\sqrt{h_1h_2P_{U1}P_{U2}}}{1+g_1P_1+g_2P_2+2\sqrt{g_1g_2P_{U1}P_{U2}}},\\
    T_2&=\frac{[1+h_{12}(P_{10}+P_{12})][1+h_{21}(P_{20}+P_{21})]}{1+g_1P_1+g_2P_2+2\sqrt{g_1g_2P_{U1}P_{U2}}},\\
\text{and} \qquad
T_3&=\frac{1+h_1P_{10}+h_2P_{20}}{(1+P_{10}h_{12})(1+P_{20}h_{21})}.
\end{align*}
Note that $P_{10}+P_{12}=P_1-P_{U1}$ and $P_{20}+P_{21}=P_2-P_{U2}$. Hence,
given $P_1$, $P_2$, $P_{U1}$ and $P_{U2}$, $T_1$ and $T_2$ are not related to
$P_{10}$ and $P_{20}$.

When $h_{12} \geq h_1$ and $h_{21} \geq h_2$, $T_3 \leq 1$ for any power
allocation pair $(P_{10},P_{20})$. Furthermore, $T_3 = 1$ can be achieved only
when $P_{10}=P_{20}=0$. Therefore, we have the desired result.

\bibliographystyle{IEEEtran}
\bibliography{MacFC}

\end{document}